\documentclass[11pt]{article}
\usepackage{moriond,epsfig}
\usepackage{latexsym}
\usepackage{amsmath}
\usepackage{amsfonts}
\usepackage{amsxtra}

\def\be{\begin{equation}}
\def\ee{\end{equation}}
\def\bea{\begin{eqnarray}}
\def\eea{\end{eqnarray}}

\newcommand{\emu}{e_{\mu}}

\newcommand{\tr}{\text{Tr}}
\newcommand{\re}{\mathbb{R}\text{e}}

\def\ms{\overline{\text{MS}}}

\begin{document}
\begin{flushright}
\begin{tabular}{l}
{\tt CPHT-PC 028.0505}
\end{tabular}
\end{flushright}
\vspace*{4cm}
\title{Gluon and Ghost Propagators in Landau Gauge on the Lattice}

\author{ A.Y.~Lokhov\footnote{Talk given by Alexey Lokhov.}, C.~Roiesnel}

\address{Centre de Physique Th\'eorique\footnote{Unit\'e Mixte de
    Recherche C7644 du CNRS} de l'Ecole Polytechnique, 91128 Palaiseau
    Cedex, France.}

\maketitle

\abstracts{ We study the ultraviolet behaviour of the ghost and gluon
  propagators in quenched QCD using lattice simulations. Extrapolation
  of the lattice data towards the continuum allows to use perturbation
  theory to extract $\Lambda_{\text{QCD}}$ - the fundamental parameter
  of the pure gauge theory. The values obtained from the ghost and
  gluon propagators are coherent. The result for pure gauge SU(3) at
  three loops is $\Lambda_{\ms}\approx 320\text{MeV}$.
  However this value does depend strongly upon the order of
  perturbation theory and upon the renormalisation description of the
  continuum propagators. Moreover, this value has been obtained
  without taking into account possible power corrections to the
  short-distance behaviour of correlation functions.  }

\section{Introduction}

Correlation functions involving ghosts appear in Schwinger-Dyson
equations in covariant gauge. Thus, these correlation functions play
an important role in the non-perturbative studies of the infrared
behaviour of the gluon propagator and running coupling constant, and
of the confinement.  Lattice simulations provide a non-perturbative
way to obtain propagators and vertices in Landau gauge.  The ghost
propagator has not been studied \cite{GHSU3} so much as the gluon propagator in the
case of QCD ($SU(3)$ gauge group). That is why one should first probe
the lattice formulation of the ghost propagator by studying its
behavior at small distances, where lattice results may be checked
using perturbation theory.  Moreover, the perturbative fit of the
ghost and gluon propagators gives us the value of
$\Lambda_{\text{QCD}}$ - the fundamental parameter of the pure gauge
theory, which is a renormalisation group invariant, but is
scheme-dependent. Even if this parameter is not a physical quantity,
it can serve as a definition of the unique energy scale of quenched
QCD. In this contribution we report on our on-going study of the ghost
and gluon propagators in the domain of energy 2\,GeV $\leftrightarrow$
6\,GeV.  We have fitted these correlation functions using three-loop
perturbative expansion in the $\widetilde{\text{MOM}}$ renormalisation
scheme \cite{Chetyrkin:2000dq}.  This allows us to obtain two values
for $\Lambda_{\text{QCD}}$, and thus to check the self-consistency of
the lattice approach.  In what follows we will describe briefly the
lattice formulation in the case of quenched QCD and present
preliminary results for two-points correlation functions.

\section{Simulation Setup}

On the lattice one works with an Euclidean formulation of QCD in
discretised four-dimensional space (with lattice spacing $a$). In
practical simulations one considers this theory in a finite volume
($V=L^4$), with periodic boundary conditions for the gluon field.
Then one can evaluate numerically the functional integrals which
define Green functions, using Monte-Carlo techniques. In our
simulations we have used the standard Wilson action to produce gauge
field configurations (more details may be found in reviews \cite{MM}).
Then the gauge has to be fixed. Presently, the Landau gauge is the
only covariant gauge which can be implemented efficiently on the
lattice. The Landau gauge is fixed by finding a minimum of the
functional
\begin{eqnarray}
F_U(\{g(x)\})=-\re\tr & \sum_{x,\mu}g(x)U_\mu(x)g^{\dagger}(x+\emu),  
\qquad g(x)\in SU(3),
\end{eqnarray}
where the link variable is related to the gauge field $A^{a}_{\mu}(x)$
by $U_\mu(x)=\exp{iA^a_\mu(x) t^a}$, $t^a$ are the generators of the
gauge group. A local minimum defines explicitly the gauge
transformation that transforms $\{U\}$ to $ \{U^{(g)}\}$ so, that
$U_\mu^{(g)}(x)$ satisfies the discretised Landau condition
\begin{eqnarray}
\frac{\delta}{\delta g(x)}F_U(\{g(x)\})=0\quad \Rightarrow \quad 
\re\tr\sum_\mu \left[ U^{(g)}_\mu(x) + {U^{(g)}}^{\dagger}_\mu(x-\emu) \right] =0.
\end{eqnarray}
Once the gauge is fixed, one can calculate numerically the gluon
two-point correlation function
\begin{eqnarray}
<A^a_\mu(x) A^b_\mu(x)> = \frac{1}{Z} \int [\mathcal{D}U] A^a_\mu(x) A^b_\mu(y) e^{-S_{\text{Wilson}}[U]},
\end{eqnarray}
where $A_\mu(x) = \frac{U_\mu(x)-U^{\dagger}_\mu(x)}{2i}$. In momentum
space, and in the continuum limit, the general parametrisation of the
gluon propagator in Landau gauge reads
\begin{eqnarray}
<\tilde{A}^a_\mu(p) \tilde{A}^b_\nu(-p)>=
\frac{  G(p^2)  } {p^2}\delta^{ab}(\delta_{\mu\nu}-\frac{p_\mu p_\nu}{p^2}).
\end{eqnarray}
Thus, lattice simulations can provide non-perturbative information about
the scalar function $G(p^2)$.

The ghost propagator in covariant gauge is defined in the continum
formulation of the QCD as the inverse of Faddeev-Popov operator
\begin{eqnarray}
M^{ab}(x,y) = [\partial \cdot D ]^{ab}(x,y).
\end{eqnarray}
where $D$ is the covariant derivative in the adjoint represnetation.
On the lattice this definition should be modified. Indeed, the
Faddeev-Popov operator on the lattice is defined by the second
derivative of the functional $F_U$ at a local minimum
\begin{eqnarray}
\frac{\delta^2}{\delta g(x) \delta g(y)}F_U\left(\{g(x)\}\right) \longrightarrow M_{\text{lattice}}^{ab}(x,y),
\end{eqnarray}
where ($f^{abc}$ are the structure constants of the gauge group)
\begin{eqnarray}
\label{FP_lattice}
\begin{split}
M_{\text{lattice}}^{ab}(x,y)  &= \sum_{\mu}\biggl\{
  G_{\mu}^{ab}(x)\bigl(\delta_{y,x+\hat{\mu}}- \delta_{y,x}\bigr) + 
  G_{\mu}^{ab}(x-\hat{\mu})\bigl(\delta_{y,x-\hat{\mu}}-\delta_{y,x}\bigr) \\
 &\qquad\quad + \frac{1}{2}\sum_{c}f^{abc}\bigl(
             \delta_{y,x+\hat{\mu}}A_{\mu}^{c}(x)
           - \delta_{y,x-\hat{\mu}}A_{\mu}^{c}(x-\hat{\mu}) \bigr)
          \biggr\}
\end{split}
\end{eqnarray}
and
\begin{eqnarray}
G_{\mu}^{ab}(x) = \frac{1}{2}\,\text{Tr}\left(\left\{t^{a},t^{b}\right\}\left(U_{\mu}(x)+U_{\mu}^{\dagger}(x)\right)\right)
\end{eqnarray}
Inverting this operator with a local source at the origin with no
zero-mode component
 $$
 \left(1-\frac{1}{V}, -\frac{1}{V} , \ldots, -\frac{1}{V}\right)
 $$
 ($V$ is the volume of the lattice), and performing a Fourier
 transformation, one can extract the scalar form factor $F(p^2)$ of
 the ghost propagator in Landau gauge for all momenta at once
\begin{eqnarray}
< c^a(p) \bar{c}^b(-p)>=
\frac{  F(p^2)  }{p^2} \delta^{ab}.
\end{eqnarray}
Let us discuss briefly the errors of the method.  Obviously, one has
the statistical error due to the Monte-Carlo calculation of the
functional integral.  The systematic error is mostly due to the
space-time discretisation. Indeed, the rotational symmetry is broken
down to the group $H_4$ - the symmetry group of the hypercube. That
means that, at most momenta, the calculated values of any scalar
quantity will split, because an orbit of the $H_4$ group is not
uniquely defined by the value of $p^2$.  Thus, one should extrapolate
all data points to the continum limit $a\rightarrow 0$. Now we have a
well developed method of performing such an extrapolation
\cite{Becirevic:1999a,Becirevic:1999b}.  All plots, presented further,
have been done using this method. Another source for systematic error
comes from the Euclidean formulation of the QCD and gauge fixing
(lattice version of Gribov copies \cite{Bakeev:2003rr} ).  But the
so-called Gribov noise is negligible in the short-distace study that
is discussed here.

\section{Results}

We have done simulations on lattices with size $\{16^4, 24^4, 32^4\}$
($1000, 500, 250$ configurations, respectively) and for the respective
values of the lattice bare couplings $\beta=\{6.0, 6.2, 6.4\}$ ($\beta
= \frac{6}{g_0^2}$, where $g_0$ is the value of the bare coupling at
the cut-off energy scale $a^{-1}$). Thus, these lattices have roughly the
same physical volume.
\begin{figure}
\begin{center}
\psfig{figure=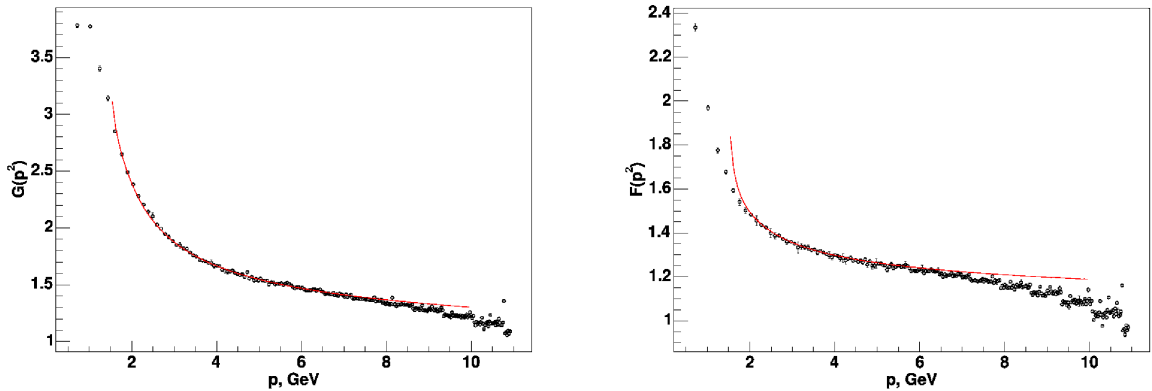, height=16.5cm,angle=-90}
\caption{ Three-loop fit for the data obtained for $V=32^2, \beta=6.4$.
The fit has been done in the interval $[2,5 \text{GeV}; 4 \text{GeV}]$ for
the gluon propagator (left image), and $[2,5 \text{GeV}; 4 \text{GeV}]$ for
the ghost propagator (right image).
\label{fig:gluon}}
\end{center}
\end{figure}
We did a fit of both propagators at three loops in the
$\widetilde{\text{MOM}}$ scheme \cite{Chetyrkin:2000dq}, which was
shown \cite{Becirevic:1999b} to provide a good description of the
lattice gluon propagator, and then converted the value of
$\Lambda_{\widetilde{MOM}}$ to the more familiar $\ms$ scheme.  The
results for $\Lambda_{\ms}$ are given in Table \ref{table_lambda}.
Two errors are quoted. The error between parentheses is the
statistical one and is rather small. The other error is the systematic
error which can be quite large, especially for the ghost propagator
where it can reach 10\%. The systematic error has been estimated from the
stability of the fits in the region $[2\text{GeV} ; 6 \text{GeV}]$.
\begin{table}[h]
\caption{Results of the fits for $\Lambda_{\text{QCD}}$}
\label{table_lambda}
\begin{center}
\begin{tabular}{|c|c|c|c|c|c|}
 \hline 
 $\beta$      & $L$  &  $\Lambda^{(3)\text{gluon}}_{\ms}$  & 
 $\chi^{2}/{\text{d.o.f}}$ & $\Lambda^{(3)\text{ghost}}_{\ms}$  & 
$\chi^{2}/{\text{d.o.f}}$   \\ \hline 
 $6.0$        &  $16$ & $324(2)^{+2~}_{-5~}\text{MeV}$  & $1.2$ 
              & $322(8)^{+20}_{-16}\text{MeV}$   & $0.5$    \\ \hline 
 $6.2$        &  $24$ & $320(2)^{+8~}_{-14}\text{MeV}$  & $0.9$ 
              & $326(5)^{+26}_{-33}\text{MeV}$  & $0.6$   \\ \hline 
 $6.4$        &  $32$ & $312(1)^{+9~}_{-25}\text{MeV}$  & $1.4$ 
              & $331(4)^{+42}_{-35} \text{MeV}$  & $0.9$ \\ \hline 
\end{tabular} 
\end{center}
\end{table}
Examples of the fits are given in Fig. \ref{fig:gluon}.

\section{Conclusion}

We obtain coherent results for $\Lambda_{\text{QCD}}$ at three loops
from the ghost and gluon propagators. This means that the lattice QCD
approach is self-consistent in the ultraviolet domain when studying
correlation functions which involve ghosts. The value for
$\Lambda_{\text{QCD}}$ that we report here is approximatively equal to
$320 \text{MeV}$ in the $\ms$ scheme. However it should be emphasized
that this value does depend strongly upon the order of perturbation
theory used in the fits. This dependence manifests itself in two ways.
First, two-loop perturbation theory is also able to reproduce the data
but with a much higher $\Lambda_{\text{QCD}}$ scale. Secondly, at any
given order, the value of $\Lambda_{\text{QCD}}$ relies very much on
the renormalization scheme used to describe the propagators in the
continuum. The value of the $\Lambda_{\ms}$ scale that we quote here
should therefore be considered as an {\em effective}
$\Lambda_{\text{QCD}}$ scale.

In order to check the asymptoticity and the coherence of
$\Lambda_{\text{QCD}}$ more precisely one should also study the
ultraviolet behaviour of the gluon and ghost propagators at four loops
in several $\widetilde{MOM}$ schemes. Possible power corrections
\cite{Boucaud:2000ey} could also make the value of the
$\Lambda_{\text{QCD}}$ scale lower. We currently work on this, and the
results will be reported elsewhere.


\section*{Acknowledgements}

We thank our colleagues of the LPT-Orsay, P.~Boucaud, J.P.~Leroy,
A.~Le~Yaouanc, J.~Micheli and O. P\`ene, for many useful discussions.


\end{document}